\documentclass{article}
\usepackage{spconf}
\usepackage{color,slashbox}
\usepackage{math}
\usepackage{enumerate}
\usepackage{theorem}
\usepackage{pifont}
\usepackage{multirow}
\usepackage{hyperref}
\usepackage{cite}
\usepackage{graphicx}
\usepackage{amsmath}
\usepackage{amsfonts}
\usepackage{amssymb}
\usepackage{amsopn}
\usepackage{xspace}
\usepackage{array}
\usepackage{epsfig}
\usepackage{algorithm}
\usepackage{algorithmic}
\usepackage{float}

\title{Phase-Optimized K-SVD for Signal Extraction from \\ Underdetermined Multichannel Sparse Mixtures\vspace{-3mm}}
\name{\vspace{-5mm}Antoine Deleforge and Walter Kellermann\thanks{The research leading to these results has received funding from the
European Union’s Seventh Framework Programme (FP7/2007-2
013) under grant agreement n${}^\circ$ 609465 (EARS project).}}
\address{University of Erlangen-Nuremberg, Germany\vspace{-4mm}}
\begin{document}
%
\maketitle
\begin{abstract}
\vspace{-1mm}
We propose a novel sparse representation for heavily underdetermined multichannel sound mixtures, \textit{i.e.}, with much more sources than microphones. The proposed approach operates in the complex Fourier domain, thus preserving spatial characteristics carried by phase differences. We derive a generalization of K-SVD which jointly estimates a dictionary capturing both spectral and spatial features, a sparse activation matrix, and all instantaneous source phases from a set of signal examples. The dictionary can then be used to extract the learned signal from a new input mixture. The method is applied to the challenging problem of ego-noise reduction for robot audition. We demonstrate its superiority relative to conventional dictionary-based techniques using recordings made in a real room.
\end{abstract}
%
%

\vspace{-4mm}
\section{Introduction}
\label{sec:intro}
\vspace{-2mm}
Most interesting signals are \textit{structured}. This is what distinguishes them from mere random noise. This structure can often be expressed in terms of \textit{sparsity} in a particular basis. More precisely, if $\Ymat=[\yvect_1,\dots,\yvect_T]$ represents $T$ signal examples, there must exist a set of $K$ atoms or a \textit{dictionary} $\Dmat=[\dvect_1,\dots,\dvect_K]$ such that each signal is a linear combination of only a few atoms, \textit{i.e.}, $\Ymat\approx\Dmat\Xmat$ where $\Xmat$ is sparse. Estimating $\Dmat$ and $\Xmat$ from $\Ymat$ is a sparse instance of \textit{matrix factorization}. In audio signal processing, it is natural to seek such a factorization in the non-negative \textit{power spectral density} (PSD) domain, since the magnitude spectra of natural sounds such as speech often feature redundancy and sparsity. This approach gave rise to a large number of methods for audio signal representation and extraction within the framework of \textit{non-negative matrix factorization} \cite{schmidt2007wind,wilson2008regularized,ozerov2010multichannel,sawada2013multichannel}. In contrast, complex spectra are usually considered uninformative and therefore not investigated. 

While single-channel signals are well represented by their PSD only, disregarding the phase comes with a substantial loss of information in multichannel signals. Indeed, phase differences carry important spatial cues. For this reason, nearly all existing NMF-based methods are limited to monaural sound processing, although recent multichannel extensions of NMF have been proposed for music signal separation\cite{ozerov2010multichannel,sawada2013multichannel}. These method do not rely on sparsity and must be tuned to a known, relatively small number of target sources.

In this paper, we propose a new sparse representation for multichannel signals in the complex Fourier domain. The key novelty is to \textit{estimate} the instantaneous phases of all involved signal spectra instead of ignoring them. The proposed decomposition may be viewed as blindly unmixing a mixture of $K\gg M$ sources where $M$ is the number of microphones. The fundamental assumption is here that each source contributes a specific complex-valued spectral component to the observation, and that the gain and the activation of each source is sparse along the time axis. We first derive a phase-optimized generalization of the well-known \textit{orthogonal matching pursuit} (OMP) method \cite{pati1993orthogonal,tropp2007signal}. This generalization allows sparse coding of a multichannel signal $\yvect_t$ given a dictionary $\Dmat$, independently of instantaneous source phases. Moreover, we show that an optimal dictionary $\Dmat$ as well as all instantaneous source phases and sparse activations can be blindly estimated from a set of signal examples $\Ymat$ only. This is achieved by deriving a phase-optimized generalization of the popular K-SVD algorithm\cite{aharon2006svd}.

The proposed representation is applied to noise reduction in the context of a humanoid robot producing self-noise ('\textit{ego-noise}') when performing motor actions \cite{loellmann2014challenges}. This problem is extremely challenging for two reasons: First, the noise signal is highly non-stationary due to fast and irregular motions and collisions. This rules out the use of conventional spectral-subtraction methods such as \cite{martin1994spectral}. Second, ego-noise signals exhibit nonzero spatial coherence between microphones \cite{loellmann2014challenges}. However, they cannot be modeled by a single point source, nor even by a small set of point sources. In the case of a walking robot, clicks generated by collisions with the floor as well as full body and microphone movements are producing sounds arriving at the microphones from many directions with unknown transfer functions. This seriously limits the usefulness of spatial filtering methods such as beamforming \cite{herbordt2003adaptive} or blind source separation \cite{buchner2004trinicon,mandel2010model}.

On the positive side, motor noise signals are strongly structured. This has been exploited using noise template databases \cite{ince2009ego,ince2011assessment}. These approaches are based on vector quantization, which can be seen as a particular instance of K-SVD\cite{aharon2006svd}. Some approaches estimate the instantaneous noise PSD using Gaussian process models \cite{furukawa2013noise} or neural networks \cite{ince2009ego}. These methods rely on synchronized motor state information, which may not be reliably available in practice.

\section{Phase-Optimized Dictionary Learning}
\label{sec:method}
\vspace{-2mm}
\subsection{Modeling Large and Sparse Multichannel Mixtures}
\label{subsec:model} 
Let $\Ymat=[\yvect_1,\dots,\yvect_T]\in\mathbb{C}^{MF\times T}$ be an observed $M-$channel spectrogram with $F$ frequency bins and $T$ time frames. We use the decomposition $\yvect_t=[\yvect\tp_{1t},\dots,\yvect\tp_{Ft}]\tp$ where $\yvect_{ft}\in\mathbb{C}^M$ is the captured $M$-channel signal at $(f,t)$. We assume that $\Ymat$ is the recording of a finite but potentially large mixture of $K$ sound sources, each emitting a specific spectral shape. One intuitive interpretation of this model in the target application of a robot recording its own noise is that the sources correspond to all possible sounds that can be emitted from the various mechanical parts of the moving robot, effectively forming a signal with an intricate spatial distribution. We denote by $\avect_{fk}\in\mathbb{C}^M$ the transfer function from source $k$ to the $M$ microphones at frequency $f$. We denote by $\phi_{ft,k}\in\mathbb{C}$ with $|\phi_{ft,k}|=1$ the instantaneous phase of source $k$ at $(f,t)$. At time $t$, we assume that each source $k$ emits a fixed magnitude spectrum $\pvect_k=[p\tp_{1k},\dots,p\tp_{Fk}]\tp\in\mathbb{R}^{+F}$ multiplied by an \textit{activation factor} (gain) $x_{kt}\ge0$.
A central assumption of our model is that the activation vector $\xvect_t=[x_{1t}\tp,\dots,x_{Kt}\tp]\tp$ is \textit{sparse}, \textit{i.e.}, only a small number $S_{\textrm{max}}\ll K$ of sources is active at time $t$, and $\xvect_t$ has at most $S_{\textrm{max}}$ nonzero elements ($\|\xvect_t\|_0\le S_{\textrm{max}}$).
For all $f$ and $t$, the mixing model reads:
\begin{equation}
\label{eq:mixing_model}
 \yvect_{ft} = \textstyle\sum_{k=1}^K \phi_{ft,k}p_{fk}\avect_{fk}x_{kt} + \evect_{ft},
\end{equation}
where $\evect_{ft}\in\mathbb{C}^M$ represents some residual noise at $(f,t)$. To simplify this expression, the transfer function and the magnitude spectrum of source $k$ at frequency $f$ are combined into a single vector $\dvect_{fk}=p_{fk}\avect_{fk}\in\mathbb{C}^M$. We denote by $\Dmat\in\mathbb{C}^{MF\times K}$ the signal's \textit{dictionary} whose columns or \textit{atoms} are the vectors $\dvect_{k}=[\dvect\tp_{1k},\dots,\dvect\tp_{Fk}]\tp\in\mathbb{C}^{MF}$. In practice, each atom $k$ can be normalized so that $\|\dvect_{k}\|_2=1$ and that each entry of $\dvect_k$ associated to the first channel is real-valued and positive.  This comes without loss of generality because the source activations and instantaneous phases compensate this normalization in equation (\ref{eq:mixing_model}).

Let $\Phimat_t=[\phivect_{t,1},\dots,\phivect_{t,K}]\in\mathbb{C}^{F\times K}$ denote the matrix of all source phases at frame $t$.
We define the \textit{phase-corrected dictionary} at frame $t$ by:
\begin{equation}
 \label{eq:POdico}
 \Dmat\{\Phimat_t\} = \left[
 \begin{array}{ccc}
  \phi_{1t,1}\dvect_{11} & \hspace{-3mm}\dots &\hspace{-3mm} \phi_{1t,K}\dvect_{1K} \\
  \vdots                 &                    &\vdots                 \\  
  \phi_{Ft,1}\dvect_{F1} & \hspace{-3mm}\dots &\hspace{-3mm} \phi_{Ft,K}\dvect_{FK} \\ 
 \end{array}
 \right]\in\mathbb{C}^{MF\times K}.
\end{equation}
The model (\ref{eq:mixing_model}) can now be rewritten as
$\yvect_t = \Dmat\{\Phimat_t\}\xvect_t + \evect_t$
where $\xvect_t$ is a sparse vector. If $\Dmat\{\Phimat_t\}$ is known, estimating $\xvect_t$ in order to minimize $\evect_t$ given $\yvect_t$ is known as a \textit{sparse coding} problem \cite{aharon2006svd,tropp2007signal}. However in the considered case of a multichannel sound mixture, this would require the prior knowledge of not only the $K$ sources' transfer functions and magnitude spectra contained in $\Dmat$, but also their instantaneous phases contained in $\Phimat_t$. The latter randomly varies over time, is hard to predict, and should therefore be estimated from the signal. This yields the following novel optimization problem, which will be referred to as \textit{phase-optimized sparse coding}:
\begin{align}
 \label{eq:PO_SC}
 &\operatorname*{argmin}_{\Phimat_t,\xvect_t} \|\yvect_t - \Dmat\{\Phimat_t\}\xvect_t\|_2 \hspace{3mm} \textrm{subject to: } \nonumber \\
 &\|\xvect_t\|_0  \le S_{\textrm{max}} \hspace{1mm} \textrm{and} \hspace{2mm}
  \forall \hspace{1mm} f,k, x_{kt}  \ge 0 \hspace{2mm}, |\phi_{ft,k}| = 1.
\end{align}
Note that due to the sparsity of $\xvect_t$, only those $\phi_{ft,k}$ for which $x_{kt}>0$ intervene in the target function. The others can be ignored, leading to a sparse matrix $\Phimat_t$. Moreover, the non-negativity constraint on $x_{kt}$ is not necessary, since for any complex pair $(x_{kt},\phivect_{t,k})$, the pair $(|x_{kt}|,\frac{x_{kt}}{|x_{kt}|}\phivect_{t,k})$ leaves the cost function unchanged. This constraint is thus relaxed in the remainder of the paper.
\vspace{-4mm}
\subsection{Phase-Optimized Orthogonal Matching-Pursuit}
\label{subsec:phiOMP} 
\vspace{-2mm}
Although finding an exact solution to sparse coding was proven to be NP-hard \cite{davis1997adaptive}, a number of efficient approximate methods have been proposed \cite{pati1993orthogonal,tropp2007signal,chen1998atomic,hoyer2002non}, among which \textit{orthogonal matching pursuit} (OMP) \cite{pati1993orthogonal,tropp2007signal} is one of the most widely used due to its simplicity and high practical efficiency. In this section, we propose an algorithm inspired by OMP that addresses the phase-optimized sparse coding problem (\ref{eq:PO_SC}). This is referred to as \textit{phase-optimized orthogonal matching pursuit} (PO-OMP) and summarized in Alg.~\ref{alg:po-omp}.

\setlength{\textfloatsep}{0pt} 

\begin{algorithm}[t!]
\caption{PO-OMP}
\label{alg:po-omp}
\textbf{Input:} Signal $\yvect_t\in\mathbb{C}^{MF}$, dictionary $\Dmat\in\mathbb{C}^{MF\times K}$, sparsity number $S_\textrm{max}$ and reconstruction threshold $\tau$.\\
\textbf{Output:} Sparse activation vector $\xvect_t\in\mathbb{R}^{+K}$ and sparse phase corrections $\Phimat_t\in\mathbb{C}^{F\times K}$ so that $\yvect_t\approx\Dmat\{\Phimat_t\}\xvect_t$.\\
\hrule
\begin{algorithmic}[1]
\STATE $\widetilde{\xvect}_t^{(0)}:=[\;]$; $\widetilde{\Phimat}_t^{(0)}:=[\;]$; $\widetilde{\Dmat}^{(0)}:=[\;]$; $\rvect_t^{(0)}:=\yvect_t$; $i:=0$;
\WHILE{$i \le S_\textrm{max}$ and $\|\rvect_t^{(i)}\|_2>\tau$}
\STATE $i := i + 1$;
\STATE $\forall f,k,\; b_{fk}:= \langle\rvect^{(i-1)}_{ft}|\dvect_{fk}\rangle$; \label{ln:po_omp_part1a}
\STATE $\forall k,\; c_k := |\sum_{f=1}^Fb_{fk}|b_{fk}|^{-1}|$;
\STATE $k(i) := \operatorname*{argmax}_{k} (c_k)$;
\STATE $\widetilde{\xvect}_t^{(i)}  := [\widetilde{\xvect}_t^{(i-1)\top}, c_{k(i)}]\tp$;
\STATE $\forall f,\; \widetilde{\phivect}_{ft}^{(i)} := [\widetilde{\phivect}_{ft}^{(i-1)} , b_{fk(i)}|b_{fk(i)}|^{-1}]$; \label{ln:po_omp_part1b}
\STATE $\widetilde{\Dmat}^{(i)}:=[\widetilde{\Dmat}^{(i-1)}, \dvect_{k(i)}]$;
\REPEAT
\STATE $\widetilde{\xvect}_t^{(i)} := (\widetilde{\Dmat}^{(i)}\{\widetilde{\Phimat}^{(i)}_t\})^{\dag}\yvect_t$; \hspace{2mm}\textit{// (${}^{\dag} = $ pseudo-inverse)} \label{ln:po_omp_part2a}
\STATE $\rvect_t^{(i)} := \yvect_t-\widetilde{\Dmat}^{(i)}\{\widetilde{\Phimat}^{(i)}_t\}\widetilde{\xvect}^{(i)}_t$;
\STATE $\forall j, f,\; \widetilde{\phi}^{(i)}_{ft,j} := \frac{\displaystyle \langle\rvect^{(i)}_{ft}|\dvect_{fk(j)}\rangle + \widetilde{\phi}^{(i)}_{ft,j}\widetilde{x}^{(i)}_{jt}}
                                                              {\displaystyle|\langle\rvect^{(i)}_{ft}|\dvect_{fk(j)}\rangle + \widetilde{\phi}^{(i)}_{ft,j}\widetilde{x}^{(i)}_{jt}|}$; \label{ln:po_omp_part2b}
\UNTIL{$\Delta(\|\rvect_t^{(i)}\|_2)<\epsilon$}
\ENDWHILE
\RETURN sparse $\xvect_t$ and $\Phimat_t$ obtained from $\widetilde{\xvect}_t^{(i)}$ and $\widetilde{\Phimat}_{t}^{(i)}$.
\end{algorithmic}
\end{algorithm}

Similarly to OMP, PO-OMP is a greedy algorithm that selects the best matching dictionary atom, indexed by $k(i)$, at each iteration $i$. This is repeated either until $i$ reaches a maximum desired sparsity number $S_\textrm{max}$ or when the cost function (\ref{eq:PO_SC}) falls below a desired reconstruction threshold $\tau$. To avoid carrying large sparse matrices, we use the variables $\widetilde{\xvect}_t^{(i)}\in\mathbb{C}^{i}$, $\widetilde{\Phimat}_{t}^{(i)}\in\mathbb{C}^{F\times i}$ and $\widetilde{\Dmat}^{(i)}\in\mathbb{C}^{MF \times i}$. They respectively correspond to $\xvect_t$, $\Phimat_{t}$ and $\Dmat$ in which only rows or columns indexed by $k(1)\dots k(i)$ are kept.
Let $\rvect_t^{(i)}=[\rvect^{(i)\top}_{1t},\dots,\rvect^{(i)\top}_{Ft}]\tp\in\mathbb{C}^{MF}$ be the residual vector at iteration $i$, \textit{i.e.}, $\rvect_t^{(i)} = \yvect_t-\widetilde{\Dmat}^{(i)}\{\widetilde{\Phimat}^{(i)}_t\}\widetilde{\xvect}^{(i)}_t$ and $\rvect_t^{(0)}=\yvect_t$.
As in OMP, each iteration of PO-OMP consists of two steps. In the first step, the dictionary atom that best approximates the current residual is found.
This requires to solve:
\begin{equation}
\label{eq:phiOMP1}
\operatorname*{argmin}_{\displaystyle k,\phivect_{t,k},x_{kt}} \|\rvect_t^{(i-1)}-\dvect_k\{\phivect_{t,k}\}x_{kt}\|_2 \hspace{1mm} \textrm{s.t.} \hspace{1mm} \forall f,\; |\phi_{ft,k}| = 1
\end{equation}
where $\dvect_k\{\phivect_{t,k}\}$ denotes the $k$-th column of $\Dmat\{\Phimat_t\}$.
Using the Lagrange multiplier method to enforce the constraints on $\phivect_{t,k}$, the solution of (\ref{eq:phiOMP1}) is obtained through lines~\ref{ln:po_omp_part1a}-\ref{ln:po_omp_part1b} of Alg.~\ref{alg:po-omp}.
In the second step, all values in $\widetilde{\xvect}_t^{(i)}$ and $\widetilde{\Phimat}_{t}^{(i)}$, including values found in previous iterations, are optimized according to the $i$ atoms selected so far. This requires to solve:

\begin{align}
\operatorname*{argmin}_{\displaystyle\widetilde{\xvect}_t^{(i)}, \widetilde{\Phimat}_{t}^{(i)}} \|\yvect_t-\widetilde{\Dmat}^{(i)}\{\widetilde{\Phimat}_{t}^{(i)}\}\widetilde{\xvect}_t^{(i)}\|_2
\hspace{1mm}\textrm{s.t.} \hspace{1mm} \forall f,j,\; |\widetilde{\phi}^{(i)}_{ft,j}| = 1. \nonumber
\end{align}
We could not find a general closed-form solution to this problem. However, it can be solved iteratively by sequentially minimizing the objective function with respect to $\widetilde{\xvect}_t^{(i)}$ and each column of $\widetilde{\Phimat}_{t}^{(i)}$ separately. Convergence is considered reached when the relative variation of the residual error $\Delta(\|\rvect_t^{(i)}\|_2)$ falls below a preset threshold $\epsilon$, \textit{e.g.}, less than $0.1\%$. The values of $\widetilde{\xvect}_t^{(i)}$ and $\widetilde{\Phimat}_{t}^{(i)}$ found in previous iterations provide a good initialization for this procedure. Closed form solutions for this sequential minimization are given in lines~\ref{ln:po_omp_part2a}-\ref{ln:po_omp_part2b} of Alg.~\ref{alg:po-omp}. In practice, the residual vector $\rvect_t^{(i)}$ is reupdated after each new estimation of $\widetilde{\phivect}^{(i)}_{t,j}$ in order to improve convergence. The overall algorithm is guaranteed to decrease the residual error $\|\rvect_t^{(i)}\|_2$ at each step. As in OMP, a local minimum may be reached due to the non-convexity of the problem. However, OMP is known to perform well if $S_\textrm{max}\ll K$, and the same was observed with PO-OMP.

\vspace{-4mm}
\subsection{Phase-Optimized K-SVD}
\label{subsec:phiKSVD}
\vspace{-2mm}
PO-OMP requires a known dictionary $\Dmat$, capturing the spectral shapes and transfer functions of the $K$ sources in the mixture. This may not be available in practice. This section addresses the challenging problem of \textit{training} such a dictionary, based on a set of examples $\Ymat=[\yvect_1,\dots,\yvect_T]\in\mathbb{C}^{MF\times T}$. More formally, we seek a solution to:
\begin{align}
 \label{eq:PO-DD}
 &\operatorname*{argmin}_{\Xmat,\Dmat,\Phimat_1,\dots,\Phimat_T} \textstyle\sum_{t=1}^T \|\yvect_t - \Dmat\{\Phimat_t\}\xvect_t\|_2^2 \hspace{3mm} \textrm{subject to: } \nonumber \\
 &\|\xvect_t\|_0  \le S_{\textrm{max}} \hspace{1mm} \textrm{and} \hspace{2mm}
 \forall \hspace{1mm} f,k, |\phi_{ft,k}| = 1.
\end{align}
This is reminiscent of a sparse dictionary learning problem \cite{elad2010sparse}, except that $\Dmat$ is corrected by $\Phimat_t$ at each $t$. Dictionary learning has been widely investigated, and the most popular method is probably K-SVD \cite{aharon2006svd}, due to its simplicity and high efficiency. Following these lines, we propose a method that solves for (\ref{eq:PO-DD}), referred to as \textit{phase-optimized K-SVD} (PO-KSVD). The corresponding algorithm is summarized in Alg.~\ref{alg:po-ksvd}.

\setlength{\textfloatsep}{1\baselineskip} 
\begin{algorithm}[t!]
\caption{PO-KSVD}
\label{alg:po-ksvd}
\textbf{Input:} Signal examples $\Ymat\in\mathbb{C}^{MF\times T}$, sparsity number $S_\textrm{max}$ and reconstruction threshold $\tau$.\\
\textbf{Output:} Matrices $\Dmat\in\mathbb{C}^{MF\times K}$, $\Xmat\in\mathbb{R}^{+K\times T}$ (sparse) and $\Phimat_1,\dots,\Phimat_t\in\mathbb{C}^{F\times K}$ (sparse) so that $\yvect_t\approx\Dmat\{\Phimat_t\}\xvect_t\;\forall t$.\\
\hrule
\begin{algorithmic}[1]
\STATE Initialize $\Dmat$ with $K$ normalized, random columns of $\Ymat$;
\REPEAT
\STATE $\forall t,\; [\xvect_t,\Phimat_t] = \operatorname{po\_omp}(\yvect_t,\Dmat,S_\textrm{max},\tau)$;
\STATE $\forall k,\; \svect_*^k=\operatorname{sparse}(\xvect_*^k)$; \textit{// Non-zero indicator of $\xvect_*^k$}
\FOR{$k = 1 \to K$}
\REPEAT
\STATE Compute $\Emat_k$; \textit{// Large-bracketed term in (\ref{PO-ksvd1})}
\STATE Obtain $\dvect_k$ and $\xvect_*^k$ from $\operatorname{svd}(\Emat_k\{\bar{\Phimat}^k_*\}_{/\{\svect_*^k\}})$;
\STATE $\forall f,t,\; \phi_{ft,k}=\frac{\displaystyle\langle\evect_{ft,k}|\dvect_{fk} x_{kt}\rangle}{\displaystyle|\langle\evect_{ft,k}|\dvect_{fk}x_{kt}\rangle|}$; \label{ln:po_ksvd1}
\UNTIL{$\Delta(\|\Emat_k\|_F)<\epsilon$}
\ENDFOR
\UNTIL{$\Delta(\Sigma_{t=1}^T\|\yvect_t-\Dmat\{\Phimat_t\}\xvect_t\|^2_2)<\epsilon$}
\RETURN matrices $\Dmat$, $\Xmat$ and $\Phimat_1,\dots,\Phimat_t$ .
\end{algorithmic}
\end{algorithm}

\begin{table*}
\centering
   \begin{tabular}{|c|c|c|c|c|c|c|c|c|c|c|}
      \cline{2-11}
      \multicolumn{1}{l|}{}  & \multicolumn{4}{c|}{Waving noise} & \multicolumn{4}{c|}{Walking noise} & \multicolumn{2}{c|}{CTS}\\
      \hline
      \multicolumn{1}{|c|}{Method used}     & SDR (dB) & SIR (dB) & PESQ & CKR & SDR (dB)& SIR (dB) & PESQ & CKR & train & test\\
      \hline
      PO-KSVD+    & {\bf2.31$\pm$4.2} & {\bf22.5$\pm$2.8} & {\bf2.09$\pm$0.3} & {\bf82.9}
                  & {\bf1.83$\pm$4.5} & {\bf22.3$\pm$3.2} & {\bf2.00$\pm$0.2} & {\bf88.4} & 9.99 & 0.59 \\       
      PO-KSVD     & 1.38$\pm$3.9 & 14.4$\pm$3.5 & 2.06$\pm$0.4 & 81.1
                  & 1.45$\pm$4.3 & 19.8$\pm$3.6 & 1.80$\pm$0.2 & 87.8 & 9.99 & 0.59 \\                 
      NMF   	  & 0.07$\pm$2.6 & 7.01$\pm$4.8 & 1.38$\pm$0.2 & 50.6
                  & 1.62$\pm$3.2 & 17.9$\pm$3.1 & 1.51$\pm$0.2 & 65.2 & 4.13 & 0.01 \\
      K-SVD       & -3.91$\pm$3.6 & -1.31$\pm$4.2 & 1.46$\pm$0.3 & 45.7
                  & 1.10$\pm$4.4 & 6.08$\pm$2.4 & 1.38$\pm$0.1 & 70.1 & 0.22 & 0.04 \\                  
      mixture     & -5.37$\pm$4.0 & -3.87$\pm$4.8 & 1.42$\pm$0.3 & 43.9 
                  & 0.76$\pm$4.2 & 4.78$\pm$2.4 & 1.33$\pm$0.1 & 67.1 & - & - \\
      \hline
   \end{tabular}
   \vspace*{-1mm}
   \caption {\label{tab:results} {\small\hspace{-1mm}Average and standard deviations (Avg$\pm$Std) of the signal-to-distortion-ratios (SDR), signal-to-interfer-ratios (SIR), PESQ measures and correct keyword recognition rates in $\%$ (CKR) over 82 target speech signals, for waving and walking noises. The last columns show the average computation times (in secs) per second of signal (CTS) for training and testing the methods using MATLAB on a conventional PC.}}
   \vspace*{-3mm}
\end{table*}

Similarly to K-SVD, PO-KSVD alternates between a sparse-coding step, \textit{i.e.}, (\ref{eq:PO_SC}) and a \textit{dictionary update} step. Since the former is solved by PO-OMP, we now focus on the latter. The key idea responsible for the efficiency of K-SVD is to sequentially update each atom and associated activations, while preserving the non-zero support of $\Xmat$ found during the sparse-coding step. Let $\xvect_*^k$ denote the $k$-th row vector of $\Xmat$ and $\svect_*^k=\operatorname{sparse}(\xvect_*^k)\in\{0;1\}^{1\times T}$ denote the binary row vector indicating the non-zero elements of $\xvect_*^k$ after PO-OMP. Let $\Phimat^k_*=\{\phi_{ft,k}\}_{f=1,t=1}^{F,T}\in\mathbb{C}^{F\times T}$ denote the sparse matrix of source $k$'s instantaneous phases. For each atom $k$, the associated optimization problem can be written:
\begin{align}
 &\operatorname*{argmin}_{\displaystyle\dvect_k,\xvect_*^k,\Phimat^k_*} \left\|
   \left(
     \Ymat - \sum_{j\ne k}(\dvect_j\xvect_*^j)\{\Phimat^j_*\}
   \right) - (\dvect_k\xvect_*^k)\{\Phimat^k_*\}
 \right\|_F \nonumber \\ 
 \label{PO-ksvd1}
 &\textrm{s.t.:} \hspace{3mm} \operatorname{sparse}(\xvect_*^k)=\svect_*^k \hspace{2mm} \textrm{and}  \hspace{2mm} \forall \hspace{1mm} f,t,\; |\phi_{ft,k}| = 1.
\end{align}
Here, $\|.\|_F$ denotes the Frobenius norm. If $\Emat_k$ denotes the matrix between large brackets, the above cost function is equal to $\|\Emat_k\{\bar{\Phimat}^k_*\} - \dvect_k\xvect_*^k\|_F$
where $\bar{\Phimat}^k_*$ denotes the complex conjugate of $\Phimat^k_*$, and $\dvect_k\xvect_*^k$ is an $MF\times T$ rank-$1$ matrix. For fixed phases $\Phimat^k_*$, the solution of $\dvect_k$ and $\xvect_*^k$ is obtained via singular value decomposition (SVD) of $\Emat_k\{\bar{\Phimat}^k_*\}_{/\{\svect_*^k\}}$ where $/\{\svect_*^k\}$ means that columns corresponding to $s_{kt}=0$ have been removed (more details in \cite{aharon2006svd}). For fixed $\dvect_k$ and $\xvect_*^k$ , the update of $\Phimat^k_*$ is closed-form, and provided at line~\ref{ln:po_ksvd1} of Alg.~\ref{alg:po-ksvd}. This sequential minimization is iterated until convergence of $\|\Emat_k\|_F$, similarly to Alg.~\ref{alg:po-omp}. As in K-SVD, the convergence of PO-KSVD relies on the ability of PO-OMP to decrease the residual error with respect to the dictionary update solution. While this is not guaranteed, it can be solved by \textit{external inference}, \textit{i.e.}, for each $t$, the output of PO-OMP is only kept if it improves the reconstruction of $\yvect_t$. Convergence is then guaranteed. 

\section{Experimental Results}
\label{sec:results} 
The commercial robot NAO of Aldebaran Robotics \cite{gouaillier2008nao} was used to gather 4-channel recordings downsampled to 16 kHz in a real room with moderate reverberation level (T60$\approx$200ms). Two one-minute recordings of NAO walking on place or repeatedly waving the right arm were used for training.
The fan of the robot was on, resulting in additional stationary ego-noise which was reduced using multichannel Wiener filtering, as described in \cite{loellmann2014challenges}. For testing, 82 recordings lasting approximately 1s each of a loudspeaker emitting speech utterances from the GRID corpus \cite{cooke2006audio} were made with the fan turned off. The loudspeaker was placed 1 meter away in front of the robot, at null elevation. These speech recordings were summed with out-of-training waving or walking sequences to generate test mixtures. Spectrograms were computed using $64$ms Hamming windows with $50\%$ overlap.

PO-KSVD was used to learn a dictionary for each of the two training signals, using several values of $K\in[5,100]$ and $S_{\textrm{max}}\in[1,5]$. Best performances were obtained with $K=40$, $S_{\textrm{max}}=3$ for waving, $K=10$, $S_{\textrm{max}}=2$ for walking. The reconstruction threshold was fixed to a low value $\tau=10^{-4}$. Once ego-noise dictionaries were trained, the ego-noise signals were estimated from test mixtures using PO-OMP. The residuals were used as desired speech output. To further improve output signals, time-frequency points for which the residual PSD was less than the estimated ego-noise PSD were set to the average background noise magnitude, while preserving the phase. This masking technique is referred to as PO-KSVD+.

PO-KSVD was compared to conventional K-SVD \cite{aharon2006svd} using the same protocol and parameters. We also compared it to NMF using the versatile implementation provided by \cite{li2013versatile}. As suggested in \cite{schmidt2007wind}, the magnitude spectra of the left microphone signal raised to the power $0.7$ were used as input. The term $\lambda\|\Xmat\|_1$ was added to the conventional NMF cost function to enforce sparsity. Several values of $\lambda\in[0,4]$ and dictionary sizes $K\in[5,40]$ were tested, and best results were obtained with $\lambda=2, K=20$ for waving and $\lambda=1, K=40$ for walking. Once a non-negative dictionary is trained, a single NMF multiplicative update can be used to estimate the ego-noise PSD from a test signal. The residuals are then used as desired magnitude spectra, while the mixture phases are preserved.

Table \ref{tab:results} summarizes the signal-to-distortion and signal-to-interfer ratios SDR and SIR \cite{vincent2006performance}, as well as the PESQ measure \cite{rix2001perceptual}, the correct keyword recognition rate\footnote{The speech recognizer \textit{pocketsphinx} \cite{huggins2006pocketsphinx} was used to recognize the keywords in the GRID corpus \cite{cooke2006audio}, as defined by the CHiME challenge \cite{vincent2013second}.} (CKR) and computational times for all methods.
Scores obtained from the unprocessed mixtures are given in the last row. PO-KSVD and PO-KSVD+ significantly outperforms conventional factorization methods in terms of all the metrics used. Sound excerpts and spectrograms are provided at \url{robot-ears.eu/po_ksvd/}.

\vspace{-4mm}
\section{Conclusion}
\label{sec:conclusion}
\vspace{-3mm}
To the best of the authors' knowledge, PO-KSVD is the first method that combines sparse factorization with instantaneous phase estimation in the complex Fourier domain. This paves the road to numerous applications in multichannel audio signal processing and beyond. Compared to traditional monaural approaches, this methods preserves and exploits spatial cues. In the future, we plan to further investigate this by adding spatial constraints to the dictionary in order to achieve underdetermined blind source separation and localization.

\bibliographystyle{IEEEbib}


\end{document}